\documentclass[useAMS,usenatbib]{mn2e}
\usepackage{graphics}
\usepackage{amsmath}
\usepackage{amssymb}

\def\aap{A\&A} 
\def\apj{ApJ} 
 
\def\apjl{ApJL} 
\def\mnras{MNRAS} 
\def\araa{ARA\&A} 
\def\aj{AJ}

\def\apjs{ApJS}

\newcommand{\Mpc}{h^{-1}\,\mathrm{Mpc}}
\newcommand{\lum}{h^{-2}\,\mathrm{erg}\,\mathrm{s}^{-1}}
\newcommand{\mass}{h^{-1}\,M_{\sun}}
\newcommand{\lumnoh}{\mathrm{erg}\,\mathrm{s}^{-1}}
\newcommand{\massnoh}{M_{\sun}}
\newcommand{\lmean}{\bar{L}_X}

\newcommand{\lbracket}{\langle L_X | N_{200} \rangle}
\newcommand{\mbracket}{\langle M_{200} | N_{200} \rangle}

\newcommand{\mmean}{\bar{M}_{200}}

\newcommand{\rtwoh}{r_{200}}

\newcommand{\rtwoo}{r_{200}^{\mathcal{N}}}
\newcommand{\sigln}{\sigma_{\ell|\nu}}

\newcommand{\sigmn}{\sigma_{\mu|\nu}}
\newcommand{\sigml}{\sigma_{\mu|\ell}}

\title[$L_X$--$M$ of Galaxy Clusters]{The $L_X$--$M$ Relation of Clusters of
  Galaxies}
\author[E. S. Rykoff et al.]{
\parbox[t]{\textwidth}{E.~S.~Rykoff,$^1$
   A.~E.~Evrard,$^2$ $^3$ $^4$
   T.~A.~McKay,$^2$ $^3$ $^4$
   M.~R.~Becker,$^2$ $^5$
   D.~E.~Johnston,$^6$
   B.~P.~Koester,$^7$ $^8$
   B.~Nord,$^2$
   E.~Rozo,$^9$
   E.~S.~Sheldon,$^{10}$
   R.~Stanek,$^3$
   R.~H.~Wechsler$^{11}$}
  \vspace*{6pt} \\ 
  $^1$TABASGO Fellow, Physics Department, University of California at Santa
  Barbara, 2233B Broida Hall, Santa Barbara, CA 93106\\
  $^2$Physics Department, University of Michigan, Ann Arbor, MI 48109\\
  $^3$Astronomy Department, University of Michigan, Ann Arbor, MI 48109\\
  $^4$Michigan Center for Theoretical Physics, Ann Arbor, MI 48109\\
  $^5$Department of Physics, The University of Chicago, Chicago, IL 60637\\
  $^6$Jet Propulsion Laboratory, 4800 Oak Grove Drive, Pasadena, CA 91109\\
  $^7$Department of Astronomy and Astrophysics, The University of Chicago,
  Chicago, IL 60637\\
  $^8$Kavli Institute for Cosmological Physics, The University of Chicago,
  Chicago, IL 60637\\
  $^9$CCAPP Fellow, The Ohio State University, Columbus, OH 43210\\
  $^{10}$Center for Cosmology and Particle Physics, Physics Department, New York
  University, New York, NY 10003\\
  $^{11}$KIPAC, Physics Dept. and SLAC, Stanford University, Stanford, CA 94305}

\begin{document}


\date{Draft: February 7, 2007}


\maketitle

\label{firstpage}

\begin{abstract}

We present a new measurement of the scaling relation between X-ray luminosity
and total mass for 17,000 galaxy clusters in the maxBCG cluster sample.
Stacking sub-samples within fixed ranges of optical richness, $N_{200}$, we
measure the mean 0.1-2.4 keV X-ray luminosity, $\langle L_X \rangle$, from the
\emph{ROSAT} All-Sky Survey.  The mean mass, $\langle M_{200} \rangle$, is
measured from weak gravitational lensing of SDSS background
galaxies~\citep{joh07}.  For $9 \le N_{200} < 200$, the data are well fit by a
power-law, $\langle L_X \rangle / 10^{42}\,\lum = (
12.6^{+1.4}_{-1.3}\,\mathrm{(stat)}\,\pm1.6\,\mathrm{(sys)} )\,( \langle
M_{200} \rangle /10^{14}\,\mass)^{1.65\pm0.13}$.  The slope agrees to within
$10\%$ with previous estimates based on X-ray selected catalogs, implying that
the covariance in $L_X$ and $N_{200}$ at fixed halo mass is not large.  The
luminosity intercept is $30\%$, or $2\sigma$, lower than determined from the
X-ray flux-limited sample of \citet{rb02}, assuming hydrostatic
equilibrium. This slight difference could arise from a combination of Malmquist
bias and/or systematic error in hydrostatic mass estimates, both of which are
expected.  The intercept agrees with that derived by \citet{sebsn06} using a
model for the statistical correspondence between clusters and halos in a WMAP3
cosmology with power spectrum normalization $\sigma_8 = 0.85$.  Similar
exercises applied to future data sets will allow constraints on the covariance
among optical and hot gas properties of clusters at fixed mass.

\end{abstract}



\section{Introduction}

Bulk properties of galaxy clusters, such as galaxy richness and velocity
dispersion, X-ray temperature and luminosity, mean weak gravitational lensing
shear and X-ray hydrostatic mass, display strong internal
correlations~\citep[for reviews, see ][]{rbn02, v05}.  Although these
correlations are  anticipated by dimensional arguments --- the ``bigger
things are bigger in all measures'' perspective --- detailed theoretical expectations are complicated by baryon evolution uncertainties.  Even with some baryon model prescribed, the stochastic nature of the halo evolution in a hierarchical
clustering framework will impose variance about the mean behavior, the scale of which is likely to be sensitive to the baryon prescription.  

Improved understanding of cluster scaling relations is desirable on both
astrophysical and cosmological grounds.  Better astrophysical
models of galaxy and hot gas evolution in halos will improve constraints on
cosmological parameters obtained from studies of cluster
samples~\citep[e.g.][]{vkfjm06,rwkme07,maer07}. 

The relation between X-ray luminosity $L_X$ and halo mass $M$ is a diagnostic
of the halo baryon fraction and the entropy structure of the intracluster gas.
From X-ray data alone, constraints on the $L_X$--$M$ relation have been
obtained using either direct or statistical arguments.  The direct approach
assumes that the intracluster gas is in hydrostatic equilibrium, so that
observations of the X-ray surface brightness, which provides $L_X$, and the
X-ray temperature profile can be combined to estimate thermal pressure
gradients and, hence, cluster binding masses.  ~\citet[][hereafter RB02]{rb02}
measured the $L_X$--$M$ relation in this manner for the X-ray flux limited
HIFLUGCS sample of roughly 100 clusters~\citep{frb01}.  The statistical
approach, on the other hand, assumes a form for the likelihood, $P(L_X | M,z)$,
which is convolved with the space density of halos in a given cosmology to
predict cluster counts as a function of X-ray flux and redshift.  This approach
was applied by \citet[][hereafter S06] {sebsn06} to the flux-limited REFLEX
survey ~\citep{bsgcv04} using halo space densities calibrated for a
$\Lambda$CDM cosmology \citep{jfwcc01, emccy02}.

A third approach to the $L_X$--$M$ relation employs weak gravitational lensing
mass estimates derived from shear patterns of background galaxies in 
deep optical imaging of cluster fields.  Recently, \citet{bskce07} and
\citet{hoe07} have performed such analysis of 11 and 20 clusters, respectively,
at redshifts near $z\sim0.25$.  These studies provide a powerful cross check
between the two methods, but provide a weaker constraint on the mass scaling as
they focus on the X-ray luminous tail of the cluster population.

In this paper, we extend the dynamic range of the third approach using a large,
uniform sample.  We present the $L_X$--$M$ relation derived from $\sim 17,000$
optically selected galaxy clusters in the maxBCG
catalog~\citep{kmawe07b-mnras}.  Binning clusters by optical richness,
$N_{200}$, we measure the mean X-ray luminosity, $\lmean \equiv \lbracket$,
using data from the \emph{ROSAT} All-Sky Survey~\citep[RASS:
][]{vabbb99-mnras}.  Details of the method are described in \citet[][henceforth
R07]{rmbej07}.  The mean mass, $\mmean \equiv \mbracket$\footnote{Here,
$M_{200}$ is the mass within a sphere of radius $\rtwoh $ that encompasses a
mean density of $200 \rho_c(z)$, with $\rho_c$ the critical density. } of each
bin is measured from analysis of the stacked weak gravitational lensing signal
from SDSS galaxy images.  Measurements of the shear are presented in
\citet{she07}, and masses are derived by \citet{joh07}.

We begin with a brief description of the input data and the methods.  We then
present measurements of $\lmean$ and $\mmean$ for richness-binned maxBCG
clusters, and compare the resultant scaling relation to the relations found by
applying direct or statistical arguments to X-ray flux-limited samples.
Throughout the paper, we assume a flat, $\Lambda$CDM cosmology with $H_0 =
100\,h\,\mathrm{km}\,\mathrm{s}^{-1}$ and $\Omega_m = 1 - \Omega_\Lambda =
0.3$.  In addition, our measurements are quoted at $z=0.25$, close to the
median redshift of the cluster catalog.  Following R07, the X-ray luminosities
are measured in the 0.1-2.4 keV observer frame, and $k$-corrected to rest frame
0.1-2.4 keV at $z=0.25$.

\section{Input Data}

Input data for this study comes from two large area surveys, the Sloan Digital
Sky Survey~\citep[SDSS: ][]{yaaaa00-mnras} and RASS.  SDSS imaging data are
used to select the clusters and to measure the weak lensing shear around 
their centers.  RASS data provide 0.1-2.4 keV X-ray
luminosities, $L_X$, measured within $\rtwoh $.  Here, we briefly describe the
maxBCG cluster catalog, binning strategies, and our methods for
calculating mean mass and X-ray luminosity.

\subsection{\label{sec:catalog}maxBCG Catalog}

The maxBCG cluster catalog provides sky locations, redshift estimates, and
richness values for the cluster sample we employ.  Details of the selection
algorithm and catalog properties are published
elsewhere~\citep{kmawe07a,kmawe07b-mnras}. In brief, maxBCG selection relies on
the observation that the galaxy population of rich clusters is dominated by
bright, red galaxies clustered tightly in color (the E/S0
ridgeline).  Since these galaxies have old, passively evolving stellar
populations, their $g-r$ color closely reflects their redshift.  The brightest such red galaxy, typically located at the peak of galaxy density, defines the cluster center.

The maxBCG catalog is approximately volume limited in the redshift range $0.1
\le z \le 0.3$, with very accurate photometric redshifts ($\delta{}z \sim
0.01$).  Studies of the maxBCG algorithm applied to mock SDSS catalogs
indicate that the completeness and purity are very high, above
$90\%$~\citep{kmawe07b-mnras, rwkme07}.  The maxBCG catalog has
been used to investigate the scaling of galaxy velocity dispersion with cluster
richness~\citep{bmkwr07} and to derive constraints on the power spectrum
normalization, $\sigma_8$, from cluster number counts~\citep{rwkme07}.

The primary richness estimator used here is $N_{200}$, defined as the number of
E/S0 ridgeline cluster members brighter than $0.4\,L_{*}$ (in $i$-band) found
within $\rtwoo$ of the cluster center~\citep{hmwas05}. Analysis
of the mean weak lensing profiles of the clusters stacked in bins of
$N_{200}$~\citep[][and \S~\ref{sec:m200}]{joh07} provides an improved measure
of the scaled apertures, $\rtwoh$.  These radii are smaller, by $\sim50\%$ on
average, than $\rtwoo$.  Our mass and luminosity estimates in this letter
consistently employ $\rtwoh$ derived from the weak lensing profiles.  However,
we remain consistent with the original catalog's richness parameter, $N_{200}$,
defined within the larger $\rtwoo$ radius.

As a cross-check, we also bin the sample using a secondary richness estimator,
$L_{200}$, the total $k$-corrected $i$-band luminosity of the cluster members.
The order parameters, $N_{200}$ and $L_{200}$, are strongly correlated,
especially for the richest clusters. We use six richness bins with $N_{200}
\geq 9$, described in Table~\ref{tab:lmn200table}, based on those used in the
weak lensing analysis of \citet{she07}.  The total number of clusters is 17335.

\subsection{Measuring $\lmean$}

We measure the mean 0.1-2.4 keV X-ray luminosity, $\lmean$, of the clusters
in each bin by stacking photons from RASS.  The typical RASS exposure time for
maxBCG clusters is $\sim\,400\,\mathrm{s}$, too short to allow significant
detections for all but the brightest individual clusters.  By binning on
$N_{200}$ or $L_{200}$ we can take advantage of the large number of maxBCG
clusters.  
The stacking method is described in
detail in R07, but there are a few differences in the analysis for this letter
which we highlight here.

We stack all the photons in a scaled aperture from clusters in a given richness
or optical luminosity bin, treating the brightest cluster galaxy selected by
the maxBCG algorithm as the center of each cluster.  In R07 we show that this
assumption for centering does not introduce a significant bias in the
calculation of $\lmean$.  For consistency with the cluster mass calculations in
\citet{joh07}, we scale all radii and luminosities to $z=0.25$.  This scale
redshift is slightly larger than used in R07, but does not significantly change
$\lmean$.  After building scaled photon maps, we construct images, radial
profiles, and spectra, and calculate an unabsorbed, rest-frame mean luminosity,
$\lmean$, for each bin. The X-ray luminosities are measured
within scaled apertures, $\rtwoh$.  These radii are smaller than those used in
the analysis of R07.  However, as $L_X$ is proportional to the gas density
squared, most of the flux originates near the dense core and the effect of this
re-scaling is modest, reducing the measured values of $\lmean$ by $\sim
10\%-15\%$.


The best-fit spectral temperatures
are used to $k$-correct the observed 0.1-2.4 keV emission to the rest frame at
$z=0.25$.  Values for the case of $N_{200}$ binning are listed in Table 1.  As
shown in R07, the data are well fit by a power law form,
\begin{equation} 
\lbracket  =  (42.1 \pm 1.7) \ ( N_{200} / 40 )^{1.88 \pm 0.06} \times 10^{42} \, \lumnoh.
\label{eq:lmean}
\end{equation} 


\begin{table}
 \begin{minipage}{80mm}
 \caption{\label{tab:lmn200table}$\lbracket$ and $\mbracket$}
 \begin{tabular}{cccc}
 \hline
 $N_{200}$ Range & $\rtwoh$ & $\lmean\,[0.1-2.4\,\mathrm{keV}]$ & $\mmean$\\


                 & ($\Mpc$) & ($10^{42}\,\lum$) & ($10^{14}\,\mass$)\\
 \hline

$9 - 11$ & 0.49 & $2.72\pm0.35$ & $0.441\pm0.080$\\
$12 - 17$ & 0.55 & $5.65\pm0.37$ & $0.600\pm0.085$ \\
$18 - 25$ & 0.64 & $12.8\pm0.98$ & $0.96\pm0.13$ \\
$26 - 40$ & 0.77 & $30.6\pm2.3$ & $1.68\pm0.23$\\
$41 - 70$ & 0.90 & $56.7\pm4.5$ & $2.52\pm0.36$\\
$71 - 188$ & 1.20 & $209\pm31$ & $5.69\pm0.88$\\

\hline
\end{tabular}
\end{minipage}
\end{table}

\subsection{\label{sec:m200}Measuring $\mmean$}

The mean projected surface density contrast, measured from the shear of faint
background galaxies as a function of projected distance from the cluster center
was calculated by \citet{she07} for the binned samples employed in this letter.
However, as maxBCG clusters are not guaranteed to be centered on the mass
density peaks of dark halos, \citet{joh07} calculate mean cluster masses using
a mass model that incorporates a degree of halo mis-centering.


Specifically, the projected surface density contrast in a given bin is modeled
as a sum of four components: i) a small scale component from the stellar mass
of the central galaxy; ii) a mean NFW~\citep{nfw97} halo; iii) a large-scale
component arising from the clustering of halos (the two-halo term); and iv) a
correction for the subset of clusters with BCG centers offset from halo
centers. The mean halo mass, $\mmean$, and its uncertainty are obtained by
marginalizing over the posterior probability distribution using a Markov Chain
Monte Carlo (MCMC) method.  The resultant masses, listed in
Table~\ref{tab:lmn200table}, scale as a power-law in richness,
\begin{equation} 
\mbracket  =  (2.1 \pm 0.3) \ ( N_{200} / 40 )^{1.28 \pm 0.04} \times 10^{14} \massnoh , 
\label{eq:mmean}
\end{equation} 
where the error in the intercept is dominated by systematic uncertainties
in the photometric redshifts of background galaxies, in the absolute shear
calibration, and in the calibration of the mis-centering model.

\section{The $\lmean$--$\mmean$ Relation}
\label{sec:lm}

\begin{figure*}
\begin{center}
\scalebox{0.38}{\rotatebox{270}{\includegraphics{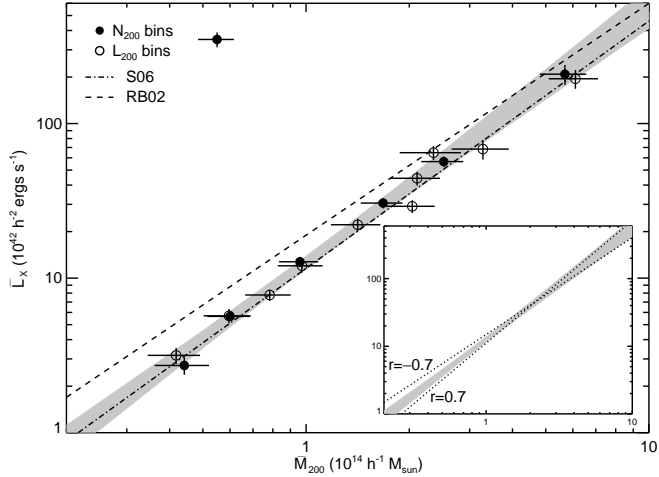}}}
\caption{\label{fig:lm}Points show maxBCG--RASS (algebraic) mean $L_X$ and
  $M_{200}$ values found by binning on $N_{200}$ (solid circles) or $L_{200}$
  (empty circles).  The dark gray band represents the $\pm 1\sigma$ contours on
  the best fit relation using the $N_{200}$ bins.  The
  dot-dashed line shows the S06 relation, while the dashed line shows the S06
  fit to the HIFLUGCS clusters from RB02 based on hydrostatic masses.  Both
  relations have been scaled assuming self-similar evolution (see
  \S~\ref{sec:lm}).  The error bar in the legend shows the typical $1\sigma$
  systematic error in the SDSS lensing masses, representing an overall shift in
  normalization that is possible in the maxBCG-RASS relation.  The inset plot
  indicates the effects on the slope due to covariance between $L_X$ and
  $N_{200}$ at fixed mass ($r$), as described in \S~\ref{sec:disc}.  If $L_X$
  and $N_{200}$ at fixed mass are correlated ($r=0.7$), this will bias the
  slope steeper, and if they are anti-correlated, this will bias the slope
  shallower.}
\end{center}
\end{figure*}

In Fig.~\ref{fig:lm} we plot the mean luminosities and lensing masses using
binning in $N_{200}$ (Table~\ref{tab:lmn200table}, filled symbols) or in
$L_{200}$ (open symbols).   The two binning choices offer consistent results.  The
gray band shows the best-fit maxBCG--RASS relation ($\pm 1\sigma$) from $N_{200}$ binning,
\begin{equation}
\lbracket = 12.6^{+1.4}_{-1.3} \left ( \frac{\mbracket}{10^{14}\,\mass} \right
)^{1.65 \pm 0.13} 10^{42}\,\lumnoh.
\label{eq:LMresult}
\end{equation}
We estimate an additional systematic error of $\pm1.6$ in normalization due to
the mass scale uncertainties described in the previous section.  The two
additional lines show results from X-ray flux-limited cluster samples.
The dashed line shows the results from the HIFLUGCS clusters of
RB02 using hydrostatic mass estimates.  The data has been re-fit by S06 after
converting their luminosity and mass estimates to $\Lambda$CDM cosmology (see
\S~3.4 in S06).  The dot-dashed line is the X-ray count-matching result of S06
for a $\Lambda$CDM model with $\Omega_m = 0.24$, spectral index $n_s=1$, and
power spectrum normalization $\sigma_8 = 0.85$; the inferred scatter in $L_X$
at fixed mass is $\sigma_{{\rm ln}L|M} = 0.40$.  Both the RB02 and S06 results
are quoted at the median redshift of the maxBCG cluster sample, $z=0.25$, by
assuming self-similar evolution, $L \sim \rho_c^{7/6}(z)$.  This raises the
$L_X$ normalization of each by 27\%.  Finally, to properly compare with the
algebraic mean luminosity of maxBCG--RASS, we adjust the normalizations of RB02
and S06 by a factor ${\rm exp}(0.5 \sigma_{{\rm ln}L|M}^2)$, resulting in an
$\sim8\%$ increase in luminosity at fixed mass.


The slope of the maxBCG--RASS relation, $1.65 \pm 0.13$, agrees well with the
values found previously.  The RB02 result, $1.50 \pm 0.08$, is somewhat
shallower, but not significantly so, and the S06 result, $1.59 \pm 0.05$, is
consistent with the present determination.  The normalization of the
maxBCG--RASS relation is consistent with the S06 results, with the RB02 masses
lying $2\sigma$ low. Further discussion of these results is
in \S~\ref{sec:disc} below.


The maxBCG--RASS normalization can be compared 
with recent lensing results published for small samples of massive
clusters at $z\sim0.25$.  \citet{bskce07} and \citet{hoe07} performed weak
lensing analyses of 11 and 20, respectively, individual clusters.  Both studies
focus on selected high luminosity clusters, with $L_X > 10^{44} \lum$, at
redshifts near the median of the maxBCG catalog.  Both papers find a
correlation between $L_X$ and $M_{200}$, but with noisy lensing mass estimates.
As each study employs a different convention for $L_X$, we have recalculated
$L_X$ for these clusters using RASS photon data (see \S~4.1 in R07).  Fixing
the slope at $1.6$, we compute the normalization of $L_X /10^{44}\lum$ at
$10^{15}\,\mass$, finding $11 \pm 2$ and $5.0 \pm 0.6$ for the \citet{bskce07} and \citet{hoe07} clusters,
respectively.  At this high mass scale, the \citet{hoe07} normalization is 
consistent with our value, while the \citet{bskce07} is $\sim2\sigma$ larger.
A full accounting for this discrepancy is beyond the scope of this work.

\section{Discussion}
\label{sec:disc}

The good level of agreement displayed in Fig.~\ref{fig:lm} among three
independent approaches to determining the $L_X$--$M$ relation indicates that
optical and X-ray selection methods are finding similar populations of massive
halos. Furthermore, the maxBCG--RASS result extends to a lower mass scale than
is probed by RB02 and S06.  In this section, we point out effects that could
lead to differences among the three measurements.  The discussion is aimed at
raising issues to be addressed by more detailed analysis in future work.

Non-zero bias in hydrostatic mass estimates, displayed in early, low-resolution
gas simulations \citep{evrard90}, is a possible source of systematic error that
would shift the RB02 result relative to the true relation.  Recent studies
using mock X-ray exposures of numerical simulations predict a systematic
underestimate of binding mass at the level of $-0.25$ in ${\rm ln}M$
\citep{rmbmd05,nvk07}.  Correcting the RB02 result by this amount would more
closely align it with the maxBCG--RASS relation.  Assuming the latter is an
unbiased estimate of the underlying halo relation, the luminosity offset
between the two relations measures the Malmquist bias arising from the X-ray
flux limit of the HIFLUGCS sample used by RB02.  Good agreement would signal a
small bias, meaning small intrinsic scatter ($\lesssim{} 10\%$) between
luminosity and mass.  Such small scatter is considered unlikely by the analysis
of S06.

A separate argument can be made based on slope estimates.  If hydrostatic mass
estimates scale with true mass as $\langle M_{\rm est} \rangle \propto M_{\rm
true}^{1+\epsilon}$, then one would expect the RB02 slope to differ by
$1.5\epsilon$ from the maxBCG--RASS value.  The measured slope difference, $0.15
\pm 0.15$, implies $\epsilon = 0.1 \pm 0.1$.  Strongly mass dependent
hydrostatic biases are therefore ruled out.


One could shift the RB02 result to higher masses without requiring a major
reduction in scatter, $\sigma_{\mathrm{ln}M|L}$, as constrained in S06.  This
would require shifting the S06 result by modifying the assumed cosmology.  The
luminosity normalization is sensitive to power spectrum normalization, $L_X
\sim \sigma_8^{-4}$, so raising $\sigma_8$ to $0.95$ would shift the S06 result
to lower $L_X$ and preserve the current level of Malmquist-bias for the RB02
result.  However, this adjustment would offset the S06 and maxBCG--RASS relations
at the $2 \sigma$ level.

Since it is binned by richness, the maxBCG--RASS result is sensitive to
covariance among $L_X$ and $N_{200}$ at fixed $M_{200}$.  Simulations suggest
mild anti-correlation, as at fixed mass high concentration halos have higher
$L_X$ but fewer galaxies~\citep{wzbka06}.  As an illustration of the effect of
covariance, consider the case of a bivariate, log-normal distribution for $L_X$
and $N_{200}$ with constant covariance.  The off-diagonal term can be
characterized by the correlation coefficient, $r \equiv \langle \delta_{{\rm
ln}L} \delta_{{\rm ln}N} \rangle$, where $\delta_{{\rm ln}X} = ({\rm ln}X- {\rm
ln}\bar{X}) / \sigma_{{\rm ln}X}$ are the normalized deviations from the mean
relation.

Consider a mass function that is a local power-law, $dn/d{\rm ln}M \sim
M^{-\alpha} = e^{-\alpha \mu}$, where $\mu \equiv \ln M$.  Convolving this
function with the bivariate log-normal, and using Bayes' theorem, allows one to
write the conditional likelihood $P(\ell , \mu | \nu) $, where $\ell \equiv \ln
L_X$ and $\nu \equiv \ln N_{200}$.  The result is a bivariate Gaussian with
mean mass $\bar{\mu}(\nu) = \bar{\mu}_0(\nu) - \alpha \sigmn^2$, with
$\bar{\mu}_0(\nu)$ the inverse of the input mean richness--mass relation and
$\sigmn$ the scatter in mass at fixed richness.  The X-ray luminosity
at fixed optical richness is distributed in a log-normal manner with mean
\begin{equation} 
\bar{\ell}(\nu) \ = \ p \cdot \left( \bar{\mu}(\nu) + \alpha \, r  \, \sigmn  \sigml \right) ,
\label{eq:lbar}
\end{equation}
and variance 
\begin{equation} 
\sigln^2 \ = \ p^2 \cdot ( \sigmn^2 + \sigml^2  - 2 r \sigmn  \sigml ), 
\label{eq:lvar}
\end{equation}
where $p$ is the slope of the halo $L_X-M_{200}$ relation.  

When $L_X$ and $N_{200}$ are independent ($r = 0$), the mean luminosity
reflects that of the mean mass selected by the richness cut.  When $r \ne 0$,
the mean is shifted by an amount that scales linearly with the mass function
slope $\alpha$.  If the scatter is constant, then this shift affects all
$N_{200}$-binned points equally, and the slope of the richness-binned
$L_X$-$M_{200}$ relation will be unbiased, $d \bar{\ell}(\nu) / d
\bar{\mu}(\nu) = p$.  However, in cold dark matter models, the slope $\alpha$
runs with mass, and this running can induce a change in slope at the level
$\Delta p / p = r \sigmn \sigml (d \alpha(\mu) / d \mu)$.

The inset of Fig.~\ref{fig:lm} explicitly demonstrates this effect for the case
$p = 1.6$, $\sigml = 0.25$ and $\sigmn = 0.5$.  We use a discrete set of halos
from the Hubble Volume simulation \citep{emccy02} that define a mass function
which is not a power-law \citep{jfwcc01}.  To each halo, we assign richness and
luminosity using a constant log-normal covariance, then bin the sample on
optical richness and compute mean luminosities and masses for each bin.
Imposing a correlation coefficient $r = \pm 0.7$ tilts the richness-binned
$L_X$--$M_{200}$ slope by $\sim 0.2$, or roughly $1.5\sigma$ given the error in
Eq.~(\ref{eq:LMresult}).

As discussed in \citet{nsre07}, covariance can both tilt the scaling relation
and modify the scatter.  The second observable quantity affected by $r$ is
the variance in $L_X$ at fixed richness $N_{200}$, Eq.~(\ref{eq:lvar}).  This
scatter was constrained in R07 to be $\sigln = 0.86\pm0.03$.  In our model, it
takes on values of $1.1$, $0.90$, and $0.72$ for $r=-0.7$, 0, and $0.7$,
respectively.  We stress that these values are sensitive to the assumed degree
of mass scatter at fixed $N_{200}$.  Our assumed value of $0.5$ is smaller than
that inferred by the spread in galaxy velocity dispersions at fixed
richness~\citep{bmkwr07} and employed by \citet{joh07} in the lensing analysis.
Using this larger scatter would result in $\sigln > 1$ for all $r \in [-1,1]$.
If the R07 measurement of $\sigln \lesssim 0.9$ is correct, then either the
\citet{bmkwr07} scatter is an overestimate or the simple log-normal model
employed here is insufficient.

Improved understanding of the variance is clearly desired.  Constraints on
$\sigln$ can be obtained by individual X-ray follow-up observations
of richness-selected maxBCG subsamples.  Such data can also generate
hydrostatic mass estimates, enabling constraints on the mass scatter
$\sigmn$.  Such a study would begin to test all the elements of the
full covariance $P(\ell,\nu|\mu)$ used to describe the underlying massive halo
population.

\section{Summary}

In this letter, we present a new measurement of the scaling between X-ray
luminosity and mass using $\sim17,000$ optically-selected clusters of galaxies.
X-ray luminosities are derived from RASS photon maps, while masses are
determined independently through modeling of the mean cluster lensing
profiles. Binning clusters by optical richness, we find that the mean $L_X$ and
$M_{200}$ values follow a power-law relation with slope $1.65 \pm 0.13$. The
consistency in slope with previous studies, including those based on
hydrostatic mass estimates, confirms that the optically-selected maxBCG catalog
selects a population of massive halos similar to those of X-ray samples.  The
consistency implies that systematic errors in hydrostatic mass are not strongly
scale-dependent.  Furthermore, we demonstrate how the slope can be affected by
covariance in $L_X$ and $N_{200}$ at fixed mass, and show that this covariance
is not large.
 
Follow-up X-ray observations of $N_{200}$ selected subsamples would provide key
information on the covariance, as expressed in Eq.~(\ref{eq:lvar}).  Data from
future deep surveys, selected via optical, X-ray and Sunyaev-Zel'dovich
signatures, will provide a much richer environment for addressing the complex
interplay between cosmology and covariant scaling relations.

{\bf Acknowledgements}

ESR and TAM are pleased to acknowledge financial support from NSF AST-0206277
and AST-0407061, and the hospitality of the MCTP.  ESR also thanks the TABASGO
foundation.  AEE acknowledges support from NSF AST-0708150.  We thank
A. Finoguenov, T. Reiprich, and H. Boehringer for helpful feedback.



\begin{thebibliography}{}

\bibitem[\protect\citeauthoryear{{Bardeau}, {Soucail}, {Kneib}, {Czoske},
  {Ebeling}, {Hudelot}, {Smail} \& {Smith}}{{Bardeau} et~al.}{2007}]{bskce07}
{Bardeau} S.,  {Soucail} G.,  {Kneib} J.-P.,  {Czoske} O.,  {Ebeling} H.,
  {Hudelot} P.,  {Smail} I.,    {Smith} G.~P.,  2007, \aap, 470, 449

\bibitem[\protect\citeauthoryear{{Becker}, {McKay}, {Koester}, {Wechsler},
  {Rozo}, {Evrard}, {Johnston}, {Sheldon}, {Annis}, {Lau}, {Nichol} \&
  {Miller}}{{Becker} et~al.}{2007}]{bmkwr07}
{Becker} M.~R.,  et~al.,  2007, \apj, 669, 905

\bibitem[\protect\citeauthoryear{{B{\"o}hringer}, {Schuecker}, {Guzzo},
  {Collins}, {Voges}, {Cruddace}, {Ortiz-Gil}, {Chincarini}, {De Grandi},
  {Edge}, {MacGillivray}, {Neumann}, {Schindler} \& {Shaver}}{{B{\"o}hringer}
  et~al.}{2004}]{bsgcv04}
{B{\"o}hringer} H.,  et~al.,  2004, \aap, 425, 367

\bibitem[\protect\citeauthoryear{{Ebeling}, {Voges}, {Bohringer}, {Edge},
  {Huchra} \& {Briel}}{{Ebeling} et~al.}{1996}]{evbeh96}
{Ebeling} H.,  {Voges} W.,  {Bohringer} H.,  {Edge} A.~C.,  {Huchra} J.~P.,
  {Briel} U.~G.,  1996, \mnras, 281, 799

\bibitem[\protect\citeauthoryear{{Evrard}}{{Evrard}}{1990}]{evrard90}
{Evrard} A.~E.,  1990, \apj, 363, 349

\bibitem[\protect\citeauthoryear{{Evrard}, {MacFarland}, {Couchman}, {Colberg},
  {Yoshida}, {White}, {Jenkins}, {Frenk}, {Pearce}, {Peacock} \&
  {Thomas}}{{Evrard} et~al.}{2002}]{emccy02}
{Evrard} A.~E.,  et~al.,  2002, \apj, 573, 7

\bibitem[\protect\citeauthoryear{{Finoguenov}, {Reiprich} \&
  {B{\"o}hringer}}{{Finoguenov} et~al.}{2001}]{frb01}
{Finoguenov} A.,  {Reiprich} T.~H.,    {B{\"o}hringer} H.,  2001, \aap, 368,
  749

\bibitem[\protect\citeauthoryear{{Gioia}, {Maccacaro}, {Schild}, {Wolter},
  {Stocke}, {Morris} \& {Henry}}{{Gioia} et~al.}{1990}]{gmsws90}
{Gioia} I.~M.,  {Maccacaro} T.,  {Schild} R.~E.,  {Wolter} A.,  {Stocke} J.~T.,
   {Morris} S.~L.,    {Henry} J.~P.,  1990, \apjs, 72, 567

\bibitem[\protect\citeauthoryear{{Hansen}, {McKay}, {Wechsler}, {Annis},
  {Sheldon} \& {Kimball}}{{Hansen} et~al.}{2005}]{hmwas05}
{Hansen} S.~M.,  {McKay} T.~A.,  {Wechsler} R.~H.,  {Annis} J.,  {Sheldon}
  E.~S.,    {Kimball} A.,  2005, \apj, 633, 122

\bibitem[\protect\citeauthoryear{{Hoekstra}}{{Hoekstra}}{2007}]{hoe07}
{Hoekstra} H.,  2007, astro-ph/0705.0358

\bibitem[\protect\citeauthoryear{{Jenkins}, {Frenk}, {White}, {Colberg},
  {Cole}, {Evrard}, {Couchman} \& {Yoshida}}{{Jenkins} et~al.}{2001}]{jfwcc01}
{Jenkins} A.,  {Frenk} C.~S.,  {White} S.~D.~M.,  {Colberg} J.~M.,  {Cole} S.,
  {Evrard} A.~E.,  {Couchman} H.~M.~P.,    {Yoshida} N.,  2001, \mnras, 321,
  372

\bibitem[\protect\citeauthoryear{{Johnston}, {Sheldon}, {Wechsler}, {Rozo},
  {Koester}, {Frieman}, {McKay}, {Evrard}, {Becker} \& {Annis}}{{Johnston}
  et~al.}{2007}]{joh07}
{Johnston} D.~E.,  et~al.,  2007, astro-ph/0709.1159

\bibitem[\protect\citeauthoryear{{Koester}, {McKay}, {Annis}, {Wechsler},
  {Evrard}, {Rozo}, {Bleem}, {Sheldon} \& {Johnston}}{{Koester}
  et~al.}{2007a}]{kmawe07a}
{Koester} B.~P.,  et~al.,  2007, \apj, 660, 221

\bibitem[\protect\citeauthoryear{{Koester}}{{Koester} et~al.}{2007b}]{kmawe07b-mnras}
{Koester} B.~P., et~al.,  2007, \apj, 660, 239

\bibitem[\protect\citeauthoryear{{Mantz}, {Allen}, {Ebeling} \&
  {Rapetti}}{{Mantz} et~al.}{2007}]{maer07}
{Mantz} A.,  {Allen} S.~W.,  {Ebeling} H.,    {Rapetti} D.,  2007, astro-ph/0709.4294

\bibitem[\protect\citeauthoryear{{Nagai}, {Vikhlinin} \& {Kravtsov}}{{Nagai}
  et~al.}{2007}]{nvk07}
{Nagai} D.,  {Vikhlinin} A.,    {Kravtsov} A.~V.,  2007, \apj, 655, 98

\bibitem[\protect\citeauthoryear{{Navarro}, {Frenk} \& {White}}{{Navarro}
  et~al.}{1997}]{nfw97}
{Navarro} J.~F.,  {Frenk} C.~S.,    {White} S.~D.~M.,  1997, \apj, 490, 493+

\bibitem[\protect\citeauthoryear{{Nord}, {Stanek}, {Rasia} \& {Evrard}}{{Nord}
  et~al.}{2007}]{nsre07}
{Nord} B.,  {Stanek} R.,  {Rasia} E.,    {Evrard} A.~E.,  2007, astro-ph/0706.2189

\bibitem[\protect\citeauthoryear{{Rasia}, {Mazzotta}, {Borgani}, {Moscardini},
  {Dolag}, {Tormen}, {Diaferio} \& {Murante}}{{Rasia} et~al.}{2005}]{rmbmd05}
{Rasia} E.,  {Mazzotta} P.,  {Borgani} S.,  {Moscardini} L.,  {Dolag} K.,
  {Tormen} G.,  {Diaferio} A.,    {Murante} G.,  2005, \apjl, 618, L1

\bibitem[\protect\citeauthoryear{{Reiprich} \& {B{\"o}hringer}}{{Reiprich} \&
  {B{\"o}hringer}}{2002}]{rb02}
{Reiprich} T.~H.,  {B{\"o}hringer} H.,  2002, \apj, 567, 716

\bibitem[\protect\citeauthoryear{{Rosati}, {Borgani} \& {Norman}}{{Rosati}
  et~al.}{2002}]{rbn02}
{Rosati} P.,  {Borgani} S.,    {Norman} C.,  2002, \araa, 40, 539

\bibitem[\protect\citeauthoryear{{Rozo}, {Wechsler}, {Koester}, {McKay},
  {Evrard}, {Johnston}, {Sheldon}, {Annis} \& {Frieman}}{{Rozo}
  et~al.}{2007}]{rwkme07}
{Rozo} E., et~al., 2007, astro-ph/0703571

\bibitem[\protect\citeauthoryear{{Rykoff}, {McKay}, {Becker}, {Evrard},
  {Johnston}, {Koester}, {Rozo}, {Sheldon} \& {Wechsler}}{{Rykoff}
  et~al.}{2007}]{rmbej07}
{Rykoff} E.~S., et~al.,
  2007, astro-ph/0709.1158

\bibitem[\protect\citeauthoryear{{Sheldon}, {Johnston}, {Scranton}, {Koester},
  {McKay}, {Oyaizu}, {Cunha}, {Lima}, {Lin}, {Frieman}, {Wechsler}, {Annis},
  {Mandelbaum}, {Bahcall} \& {Fukugita}}{{Sheldon} et~al.}{2007}]{she07}
{Sheldon} E.~S.,  et~al.,  2007, astro-ph/0709.1153

\bibitem[\protect\citeauthoryear{{Stanek}, {Evrard}, {B{\"o}hringer},
  {Schuecker} \& {Nord}}{{Stanek} et~al.}{2006}]{sebsn06}
{Stanek} R.,  {Evrard} A.~E.,  {B{\"o}hringer} H.,  {Schuecker} P.,    {Nord}
  B.,  2006, \apj, 648, 956

\bibitem[\protect\citeauthoryear{{Vikhlinin}, {Kravtsov}, {Forman}, {Jones},
  {Markevitch}, {Murray} \& {Van Speybroeck}}{{Vikhlinin}
  et~al.}{2006}]{vkfjm06}
{Vikhlinin} A.,  {Kravtsov} A.,  {Forman} W.,  {Jones} C.,  {Markevitch} M.,
  {Murray} S.~S.,    {Van Speybroeck} L.,  2006, \apj, 640, 691

\bibitem[\protect\citeauthoryear{{Voges}}{{Voges}}{1999}]{vabbb99-mnras}
{Voges} W., et~al.,  1999, \aap, 349, 389

\bibitem[\protect\citeauthoryear{{Voit}}{{Voit}}{2005}]{v05}
{Voit} G.~M.,  2005, Reviews of Modern Physics, 77, 207

\bibitem[\protect\citeauthoryear{{Wechsler}, {Zentner}, {Bullock}, {Kravtsov}
  \& {Allgood}}{{Wechsler} et~al.}{2006}]{wzbka06}
{Wechsler} R.~H.,  {Zentner} A.~R.,  {Bullock} J.~S.,  {Kravtsov} A.~V.,
  {Allgood} B.,  2006, \apj, 652, 71

\bibitem[\protect\citeauthoryear{{York}}{{York}}{2000}]{yaaaa00-mnras}
{York} D.~G., et~al.,  2000, \aj, 120, 1579

\end{thebibliography}
\newcommand{\noopsort}[1]{} \newcommand{\printfirst}[2]{#1}
  \newcommand{\singleletter}[1]{#1} \newcommand{\switchargs}[2]{#2#1}

\label{lastpage}

\end{document}